\documentclass[twocolumn,showpacs,preprintnumbers,amsmath,amssymb]{revtex4}
\usepackage{graphicx}
\usepackage{dcolumn}
\usepackage{bm}
\usepackage{color}
\usepackage{epstopdf}
\begin{document}
\newcommand{\kvec}{\mbox{{\scriptsize {\bf k}}}}
\newcommand{\lvec}{\mbox{{\scriptsize {\bf l}}}}
\newcommand{\qvec}{\mbox{{\scriptsize {\bf q}}}}
%
\def\eq#1{(\ref{#1})}
\def\fig#1{\hspace{1mm}Fig. \ref{#1}}
\def\tab#1{\hspace{1mm}Tab. \ref{#1}}
%
\preprint{Cite this article as: \textcolor{blue}{A.P. Durajski, Front. Phys. \textbf{11}, 117408 (2016)}}
\title{
---------------------------------------------------------------------------------------------------------------\\
Anisotropic evolution of energy gap in Bi2212 superconductor}
\author{A.P. Durajski}
\email{adurajski@wip.pcz.pl}
\affiliation{Institute of Physics, Cz{\c{e}}stochowa University of Technology, Ave. Armii Krajowej 19, 42-200 Cz{\c{e}}stochowa, Poland}
\date{\today}
\begin{abstract}
We present a systematic analysis of the energy gap in underdoped Bi2212 superconductor as a function of temperature and hole doping level. Within the framework of the theoretical model containing the electron-phonon and electron-electron-phonon pairing mechanism, we reproduced the measurement results of modern ARPES experiments with very high accuracy.
We showed that the energy-gap amplitude is very weakly dependent on the temperature but clearly dependent on the level of doping. The evidence for a non-zero energy gap above the critical temperature, referred to as a pseudogap, was also obtained.
\end{abstract}
\maketitle
\section{INTRODUCTION}
The existence of a pseudogap phase is one of the most interesting phenomena found in copper-oxide superconductors and has been the focus of experimental and theoretical studies for more than a quarter of a century. The origin of the pseudogap is still debated, and different experiments led to seemingly contradictory results \cite{Timusk, Chen2014}. According to one perspective, the pseudogap is interpreted as a precursor of superconducting gap in the normal state \cite{Lu2010, Tahir-Kheli}. From an alternative perspective, the pseudogap arises from some instability unrelated to the pairing process \cite{Parker2010, Lee}.

Various theories have been proposed for the mechanism of high-temperature superconductivity, but thus far, a consensus in this matter has been lacking \cite{GZhao, Keimer, Eduardo, SzczesniakPLA, SzczesniakSSC, Kumala2005, Gladysiewicz, Gonczarek, Choi2012, Tarasewicz1, Tarasewicz2}.
Recent attempts to describe the thermodynamic properties of the superconducting state in cuprates within the framework of the microscopic theory based on the electron-phonon (EPh) and electron-electron-phonon (EEPh) interactions has been presented and tested in various papers \cite{Szczesniak2012a, Durajski2014a, SzczesniakActa, SzczesniakJarosik}.
The introduced theory has two main input parameters: the effective pairing potential for the EPh interaction (V) and the EEPh interaction (U). 
The first parameter can be determined based on the value of the critical temperature ($T_C$), while the second one depends on $T_C$ and the pseudogap temperature ($T^{\star}$) and is chosen such that it reproduces the value of $T^{\star}$.
Here, above the critical temperature, the superconducting state disappears, and below the pseudogap temperature, in the normal state, a pseudogap appears in the electron density of states.
At this point, it should be noted that the potential V is a unique function of the critical temperature because the Hamiltonian, from which we started, has been obtained from a partial canonical transformation (details can be found in the paper \cite{Szczesniak2012a}). In fact, both V and U should depend on $T_C$ and $T^{\star}$. The full equation for the order parameter can be found in another paper \cite{Szczesniak2015A}, the results of which are illustrated to be qualitatively consistent with the results of the simplified model presented here (only a rescaling of V and U values is necessary).

In the present paper, within the framework of the above mentioned model, a quantitative analysis of the average amplitude of the energy gap as a function of temperature and hole doping level was performed for Bi$_2$Sr$_2$CaCu$_2$O$_{ 8+\delta}$ (Bi2212) superconductor.
The obtained results were compared with the results of recent high-quality angle-resolved photoemission (ARPES) experiments on this system \cite{Vishik2010, Vishik2012}.
\section{Theoretical model}
The most general form of the Hamiltonian that contains the essential physics of the pairing mechanism for cuprates is as follows \cite{Szczesniak2012a}:
\begin{equation}
\label{r-1}
H\equiv H^{\left(0\right)}+H^{\left(1\right)}+H^{\left(2\right)},
\end{equation}
where 
\begin{equation}
\label{r-2}
H^{\left(0\right)}\equiv\sum_{\kvec\sigma }\varepsilon _{\kvec}c_{\kvec\sigma
}^{\dagger}c_{\kvec\sigma }+\sum_{\qvec}\omega _{\qvec}b_{\qvec}^{\dagger}b_{\qvec},
\end{equation}
\begin{equation}
\label{r-3}
H^{\left(1\right)}\equiv\sum_{\kvec\qvec\sigma }g^{\left(1\right)}_{\kvec}\left({\bf q}\right)
c_{\kvec+\qvec\sigma}^{\dagger}c_{\kvec\sigma}\varphi_{\qvec},
\end{equation}
and
\begin{equation}
\label{r-4}
H^{\left(2\right)}\equiv\sum_{\kvec\kvec^{'}\qvec\lvec\sigma}
g^{\left(2\right)}_{\kvec,\kvec^{'}}\left({\bf q},{\bf l}\right)
c_{\kvec-\lvec\sigma }^{\dagger}c_{\kvec\sigma}
c_{\kvec^{'}+\lvec+\qvec-\sigma}^{\dagger}c_{\kvec^{'}-\sigma}\varphi_{\qvec}.
\end{equation}
The non-interacting electrons and phonons are described by $H^{\left(0\right)}$, where the band energy for the two-dimensional square lattice can be expressed as $\varepsilon _{\kvec}=-t\gamma\left({\bf k}\right)$. Symbol $t$ denotes the hopping integral, and $\gamma\left({\bf k}\right)\equiv 2\left[\cos\left(k_{x}\right)+\cos\left(k_{y}\right)\right]$. 
In the case of Bi2212, we assume that
\begin{equation}
\label{eqt}
t= \begin{cases} ~~350~{\rm meV}  &\text{for the nearest-neighbours \cite{Tohyama,Kim}}
\\~~0 &\text{in other cases.} \end{cases} \nonumber
\end{equation}

Symbols $c^{\dagger}_{\kvec\sigma}$ and $c_{\kvec\sigma}$ denote the creation and annihilation operators, respectively, for the electron with momentum ${\bf k}$ and spin $\sigma$. Function $\omega_{\qvec}$ models the energy of the phonon with wave number ${\bf q}$. Operators $b^{\dagger}_{\qvec}$ and $b_{\qvec}$ are the phonon creation and annihilation operators, respectively.

The EPh and EEPh terms are given by $H^{\left(1\right)}$ and $H^{\left(2\right)}$, where: $\varphi_{\qvec}\equiv b_{-\qvec}^{\dagger}+b_{\qvec}$. Symbol $g^{\left(1\right)}_{\kvec}\left({\bf q}\right)\simeq g^{\left(1\right)}$ denotes EPh coupling \cite{Frohlich1954A} and 
$g^{\left(2\right)}_{\kvec,\kvec^{'}}\left({\bf q},{\bf l}\right)\simeq g^{\left(2\right)}$ models the EEPh interaction \cite{Szczesniak2012a}.

Within the framework of the abovementioned model, the dependence of the energy gap on the momentum should be calculated from the anomalous thermodynamic average \cite{Durajski2014a, Durajski2014Hg}:
\begin{eqnarray}
\label{r4}
\phi_{\kvec}&=&
\left(\frac{1}{N_{0}}\sum^{\omega_{0}}_{\kvec_{1}}\eta\left(\bf k_{1}\right)\phi_{\kvec_{1}}\right)\\ \nonumber
&\times&
\left[V+U\left(\frac{1}{N_{0}}\sum^{\omega_{0}}_{\kvec_{2}}\eta\left(\bf k_{2}\right)\phi_{\kvec_{2}}\right)
\left(\frac{1}{N_{0}}\sum^{\omega_{0}}_{\kvec_{3}}\eta\left(\bf k_{3}\right)\phi^{\star}_{\kvec_{3}}\right)\right]\\ \nonumber
&\times&
\eta\left(\bf {k}\right)\chi_{\kvec},
\end{eqnarray}
where
\begin{equation}
\label{r5}
\chi_{\kvec}\equiv\frac{\tan\left[\frac{i\beta}{2}\sqrt{\varepsilon^{2}_{\kvec}+M^{2}_{\kvec}}\right]}
{2i\sqrt{\varepsilon^{2}_{\kvec}+M^{2}_{\kvec}}},
\end{equation}
and
\begin{eqnarray}
\label{r6}
M^{2}_{\kvec}&\equiv&
\eta^{2}\left({\bf k}\right)\left(\frac{1}{N_{0}}\sum^{\omega_{0}}_{\kvec_{1}}\eta_{\kvec_{1}}\phi^{\star}_{\kvec_{1}}\right)
\left(\frac{1}{N_{0}}\sum^{\omega_{0}}_{\kvec_{2}}\eta_{\kvec_{2}}\phi_{\kvec_{2}}\right)\\ \nonumber
&\times&
\left[V+U\left(\frac{1}{N_{0}}\sum^{\omega_{0}}_{\kvec_{3}}\eta_{\kvec_{3}}\phi_{\kvec_{3}}\right)
\left(\frac{1}{N_{0}}\sum^{\omega_{0}}_{\kvec_{4}}\eta_{\kvec_{4}}\phi^{\star}_{\kvec_{4}}\right)\right]^{2}.
\end{eqnarray}
Finally, the energy gap is defined as follows: 
\begin{equation}
\label{r7}
G_{\kvec}(T)\equiv 2\eta\left(\bf{k}\right)\left|\phi_{\kvec}\right|\left[V+U\left|\eta\left(\bf{k}\right)\right|\left|\phi_{\kvec}\right|^{2}\right],
\end{equation}
where $\eta\left({\bf k}\right)\equiv 2\left[\cos\left(k_{x}\right)-\cos\left(k_{y}\right)\right]$ introduces a $d_{x^2 - y^2}$ ($d$-wave) symmetry.
The anomalous thermodynamic average has been solved for $7000$ points close to the Fermi energy.

In this approach, finite doping is included in the values of the pairing potentials $V$ and $U$ because $T_C$ and $T^{\star}$ are functions of doping. The presented analysis method allows the partial simulation of the influence of the chemical potential on the energy gap. It should be noted that since the correlated band is narrow and practically all electrons participate in pairing, the equation for the chemical potential should be possess. Therefore, extremely complex numerical calculations should be performed, for example, in the framework of the Eliashberg formalism, where the summation by \textbf{k} and the Matsubara frequencies are taken into consideration.
However, even within the framework of presented simplified model, we can obtain a very good agreement between computational and experimental results.

\section{Results}

\begin{figure}[h]
\includegraphics*[width=\columnwidth]{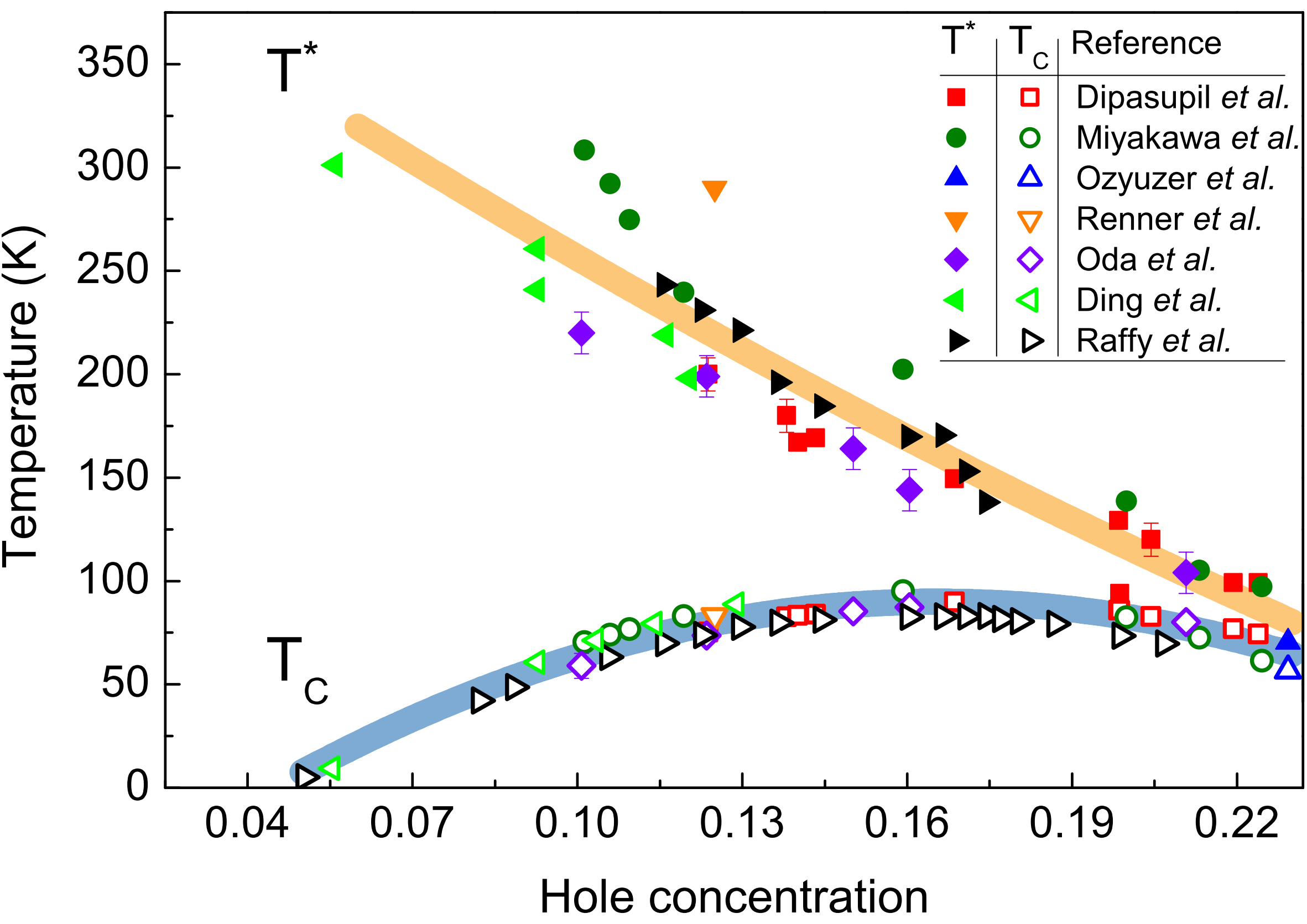}
\caption{$T_{C}$ and $T^{\star}$ as a function of hole concentration. The experimental results are taken from: 
Dipasupil {\it et al.} \cite{Dipasupil}, 
Miyakawa {\it et al.} \cite{Miyakawa}, 
Ozyuzer {\it et al.} \cite{Ozyuzer}, 
Renner {\it et al.} \cite{RennerBi2212}, 
Oda {\it et al.} \cite{Oda}, 
Ding {\it et al.} \cite{Ding}, and  
Raffy {\it et al.} \cite{Raffy}.}
\label{f1}
\end{figure}

The pairing potentials $V$ and $U$ should be calculated on the basis of the experimental dependencies of the critical temperature and pseudogap temperature on the hole concentration ($p$). The corresponding dependencies are shown in \fig{f1}. It should be remembered that $V$ is a unique function of the critical temperature, while $U$ depends on both $T_{C}$ and $T^{\star}$ \cite{Szczesniak2012a}. In the present study, we take into account a wide range of the hole concentrations: $p\in\left\lbrace 0.08, 0.10, 0.12, 0.16\right\rbrace $.
\begin{table}[b]
\caption{\label{t1} Pairing potentials $V$ and $U$ calculated from the experimental values of $T_{C}$ and $T^{\star}$.}
\begin{ruledtabular}
\begin{tabular}{lllll}
$p$ &$T_{C}$ (K)&$T^{\star}$ (K)&$V$ (meV)&$U$ (meV)\\
         &    		 &       &    			 &      \\\hline
$0.08	$&	$	50	$&$	277	$&$	{5.29}	$&$	{12.57}$\\
$0.10	$&	$	65	$&$	250	$&$	{5.91}	$&$	{11.24}$\\
$0.12	$&	$	75	$&$	226	$&$	{6.30}	$&$	{10.19}$\\
$0.16	$&	$	92	$&$	166	$&$	{6.93}	$&$	{7.49}$\\
\end{tabular}
\end{ruledtabular}
\end{table}
\begin{figure*}[!t]
\includegraphics[width=0.5\columnwidth]{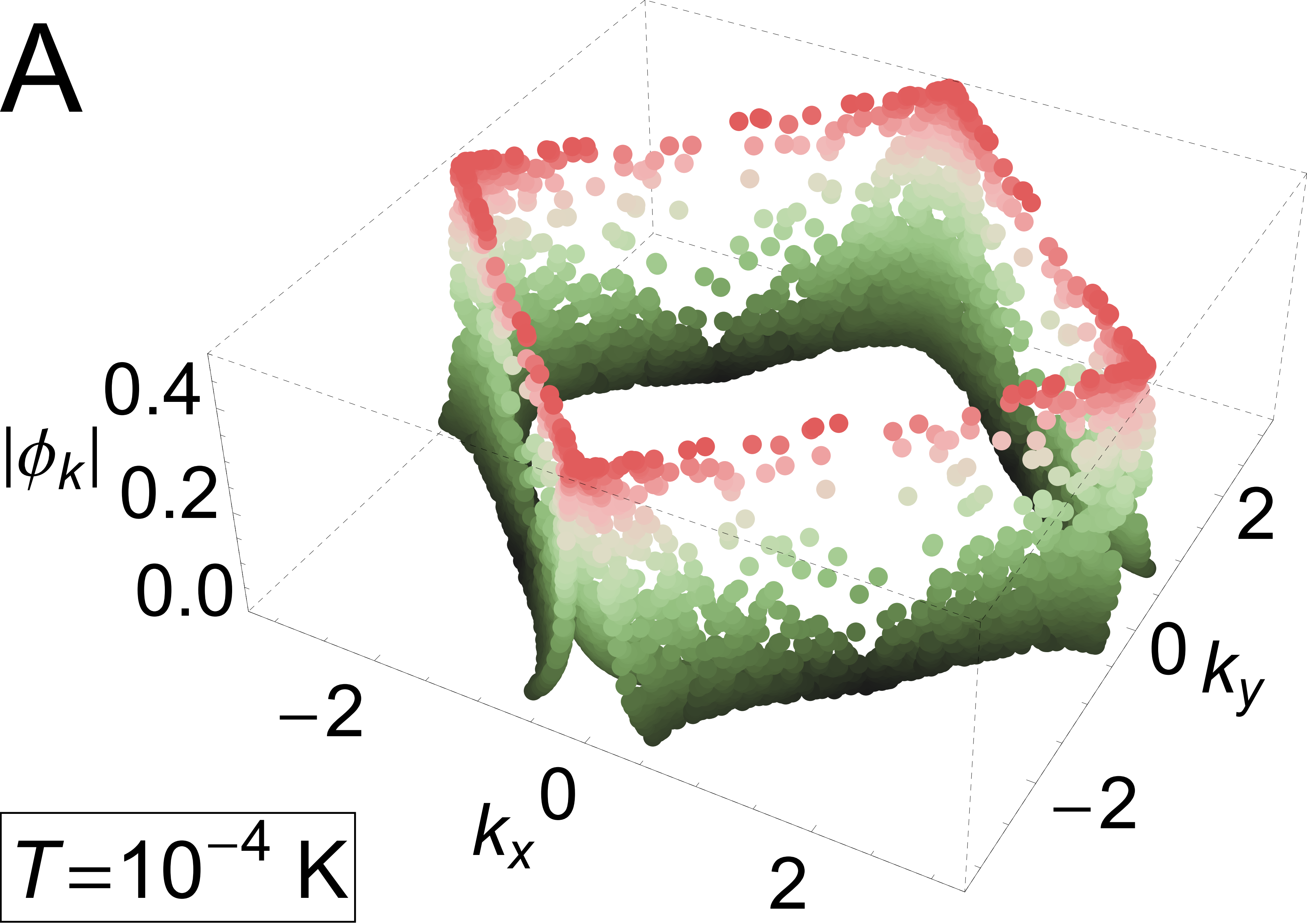}
\includegraphics[width=0.5\columnwidth]{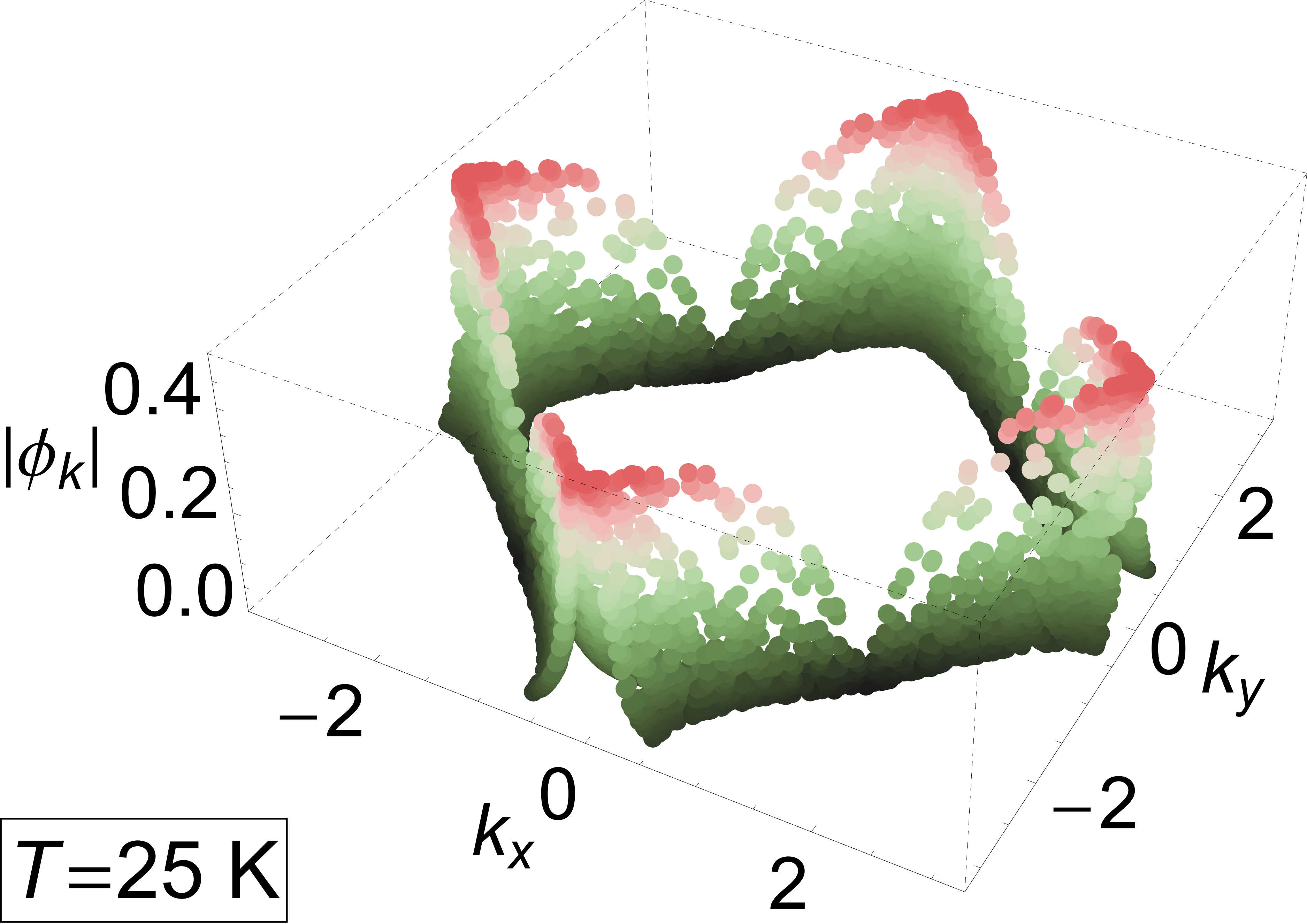}
\includegraphics[width=0.5\columnwidth]{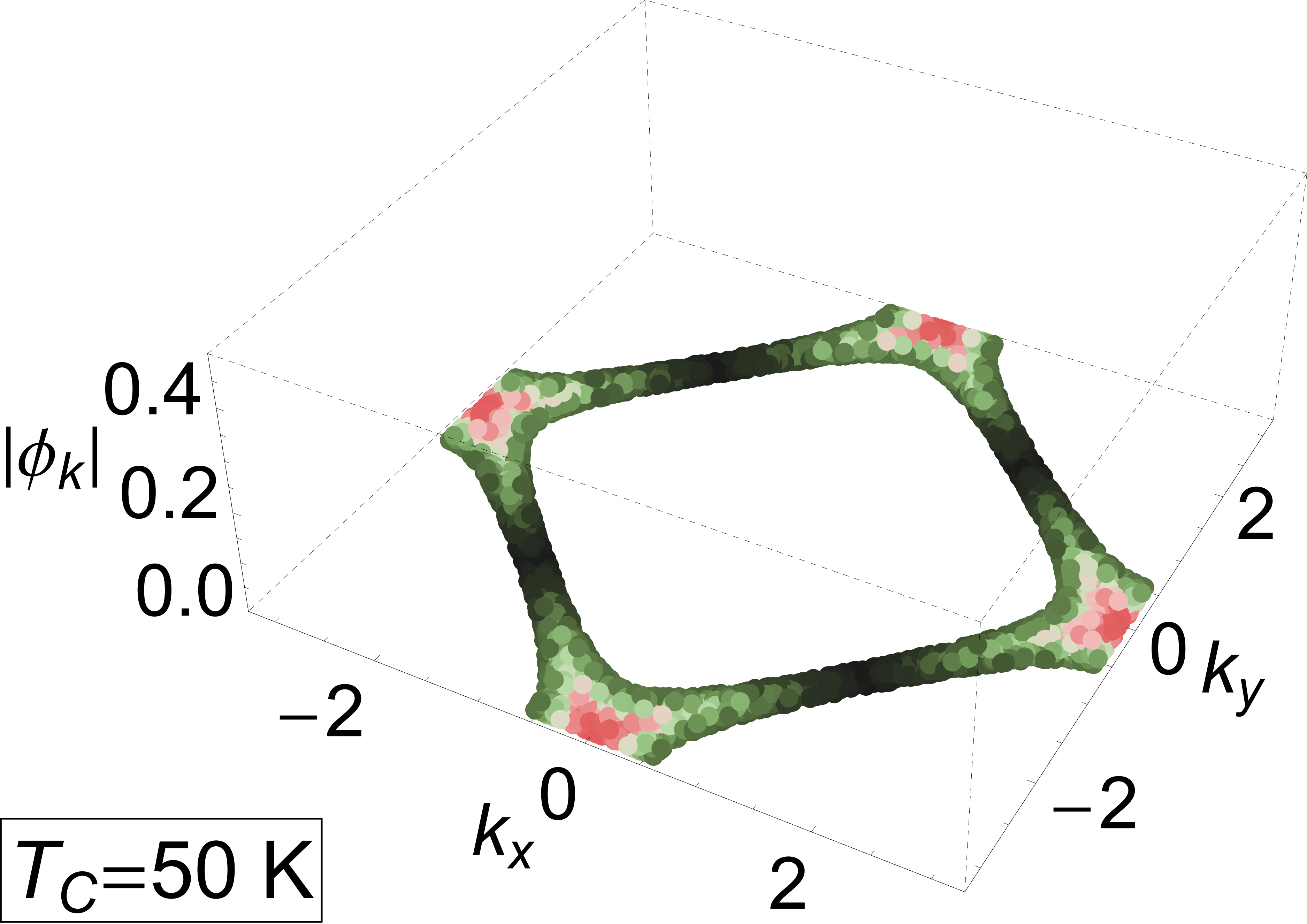}
\includegraphics[width=0.5\columnwidth]{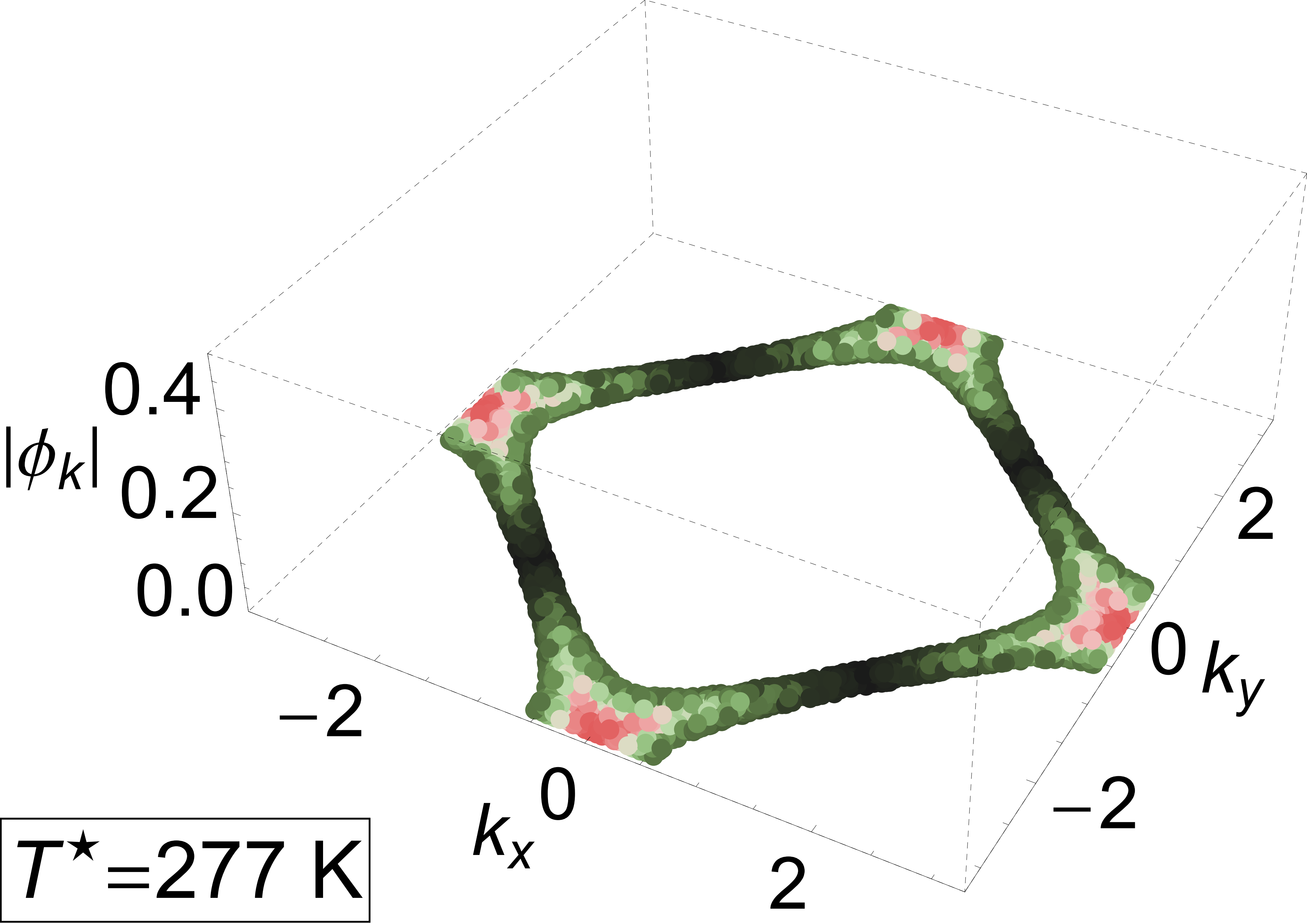}\\\vspace{0.5cm}
\includegraphics[width=0.5\columnwidth]{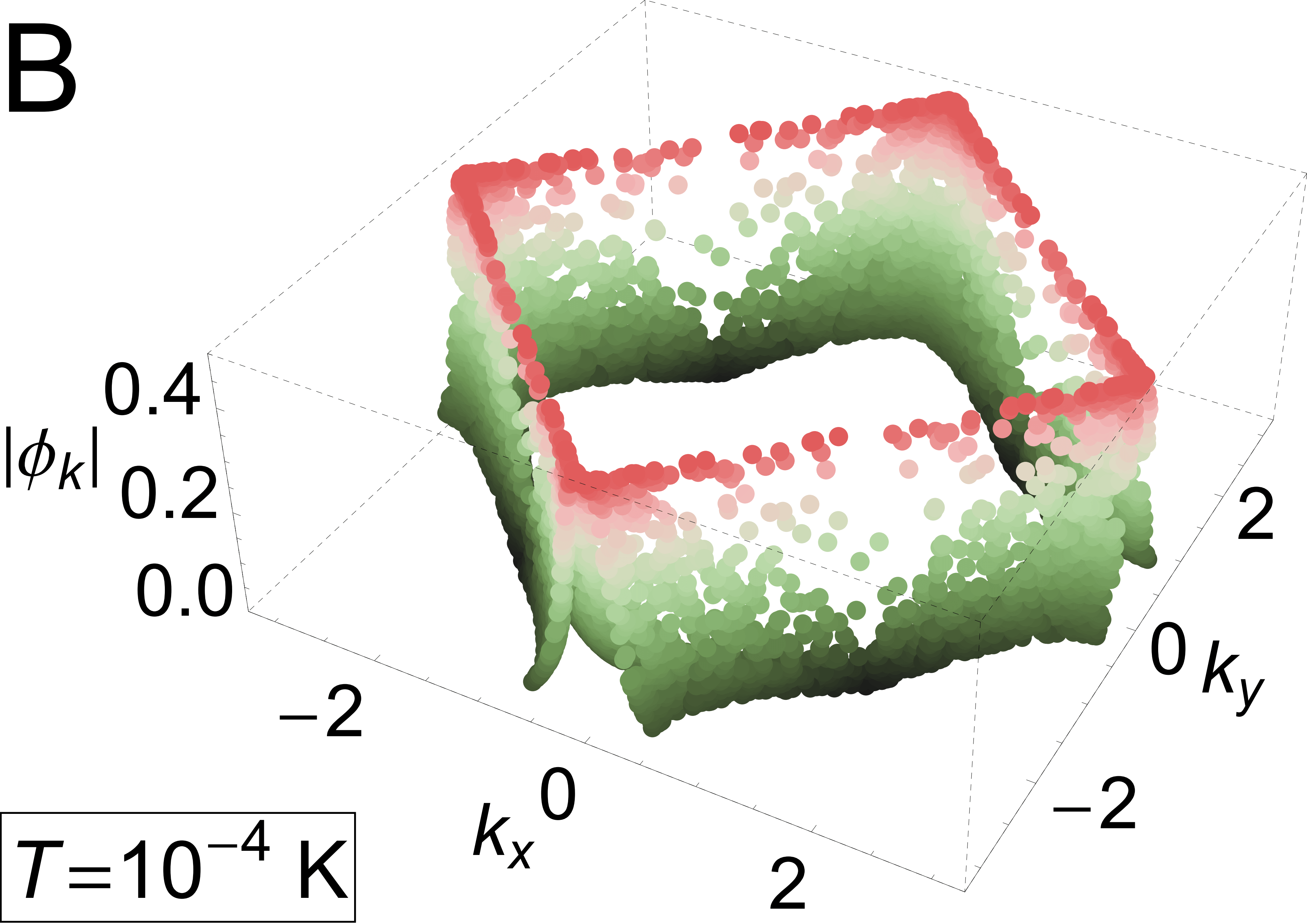}
\includegraphics[width=0.5\columnwidth]{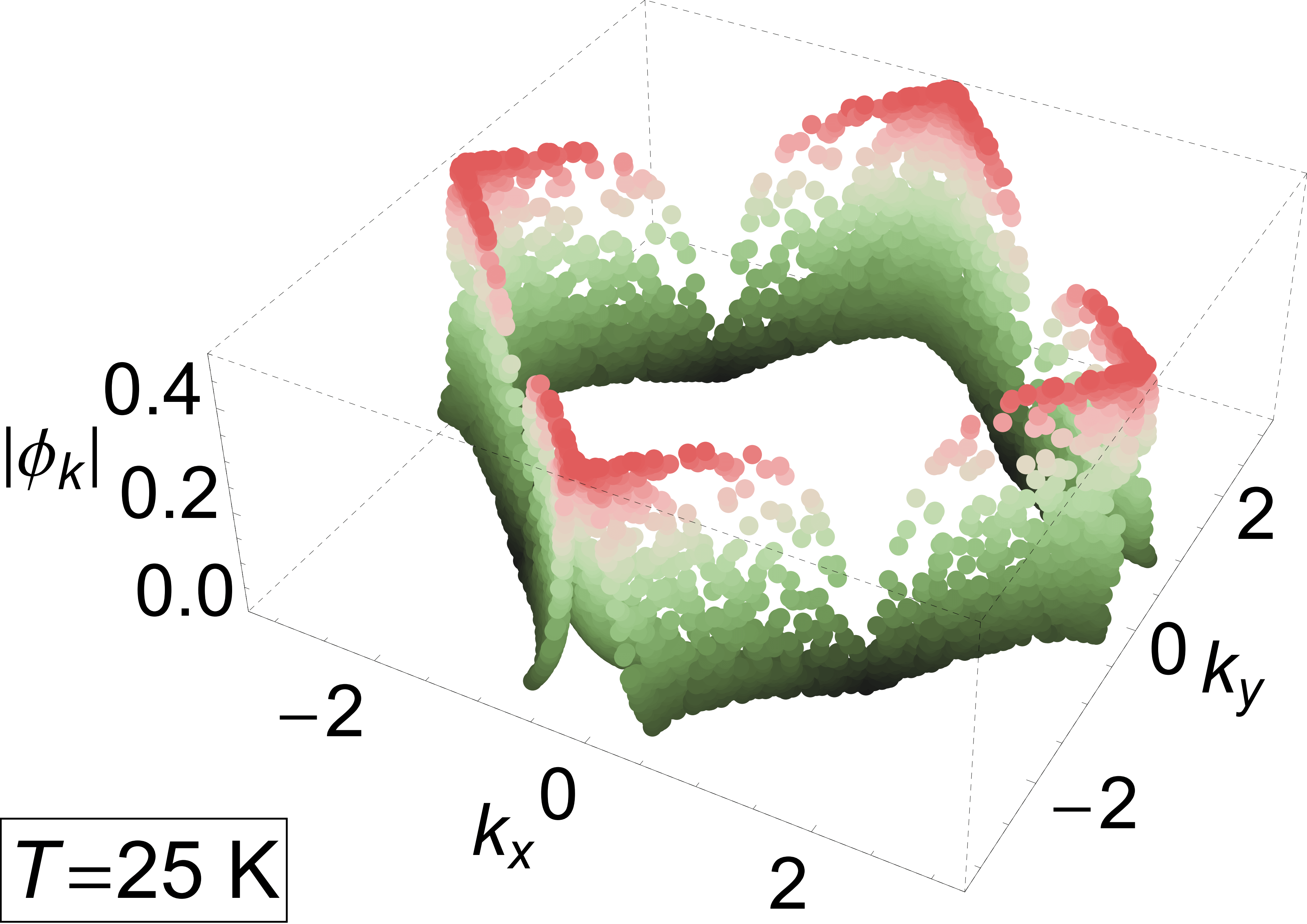}
\includegraphics[width=0.5\columnwidth]{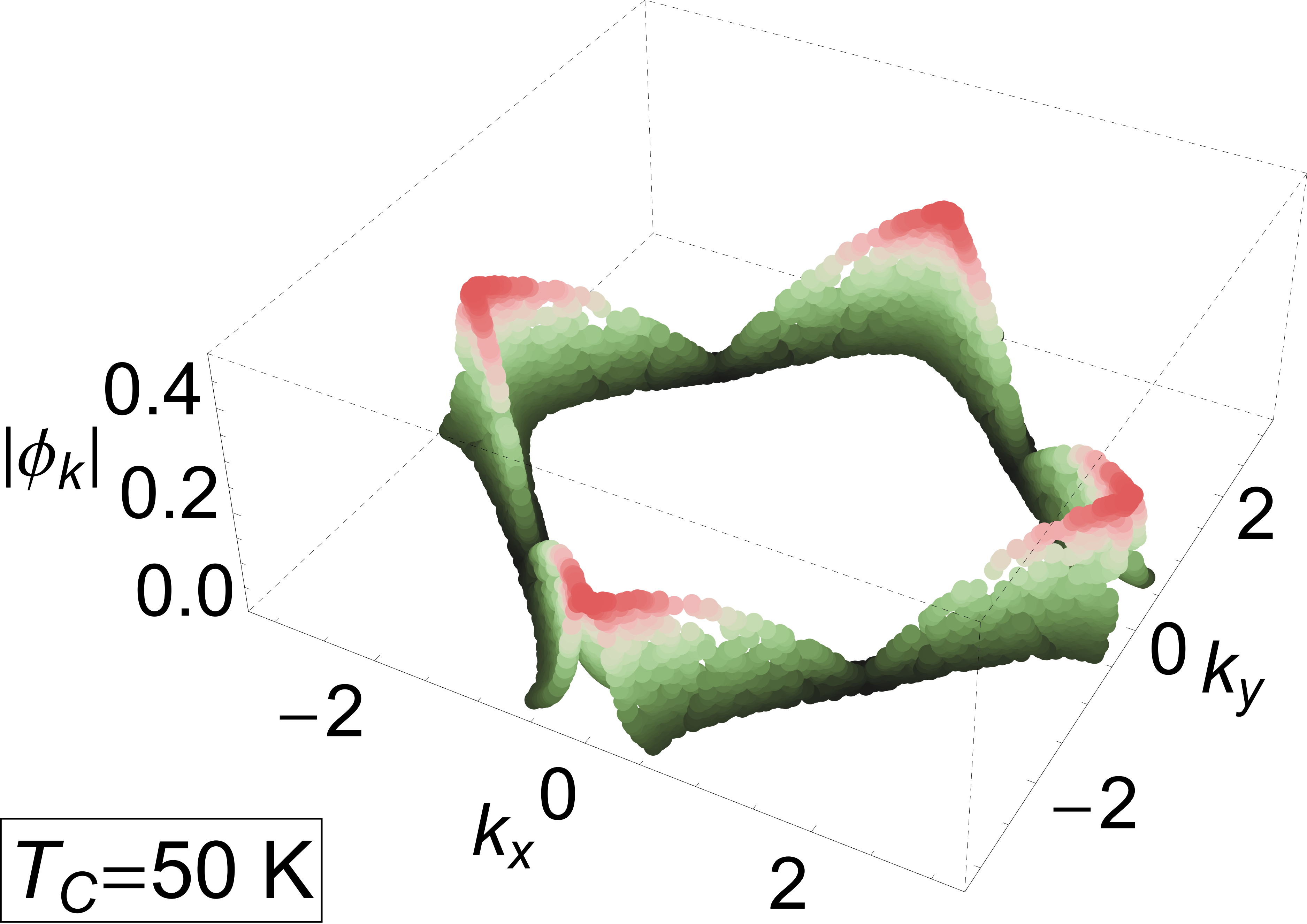}
\includegraphics[width=0.5\columnwidth]{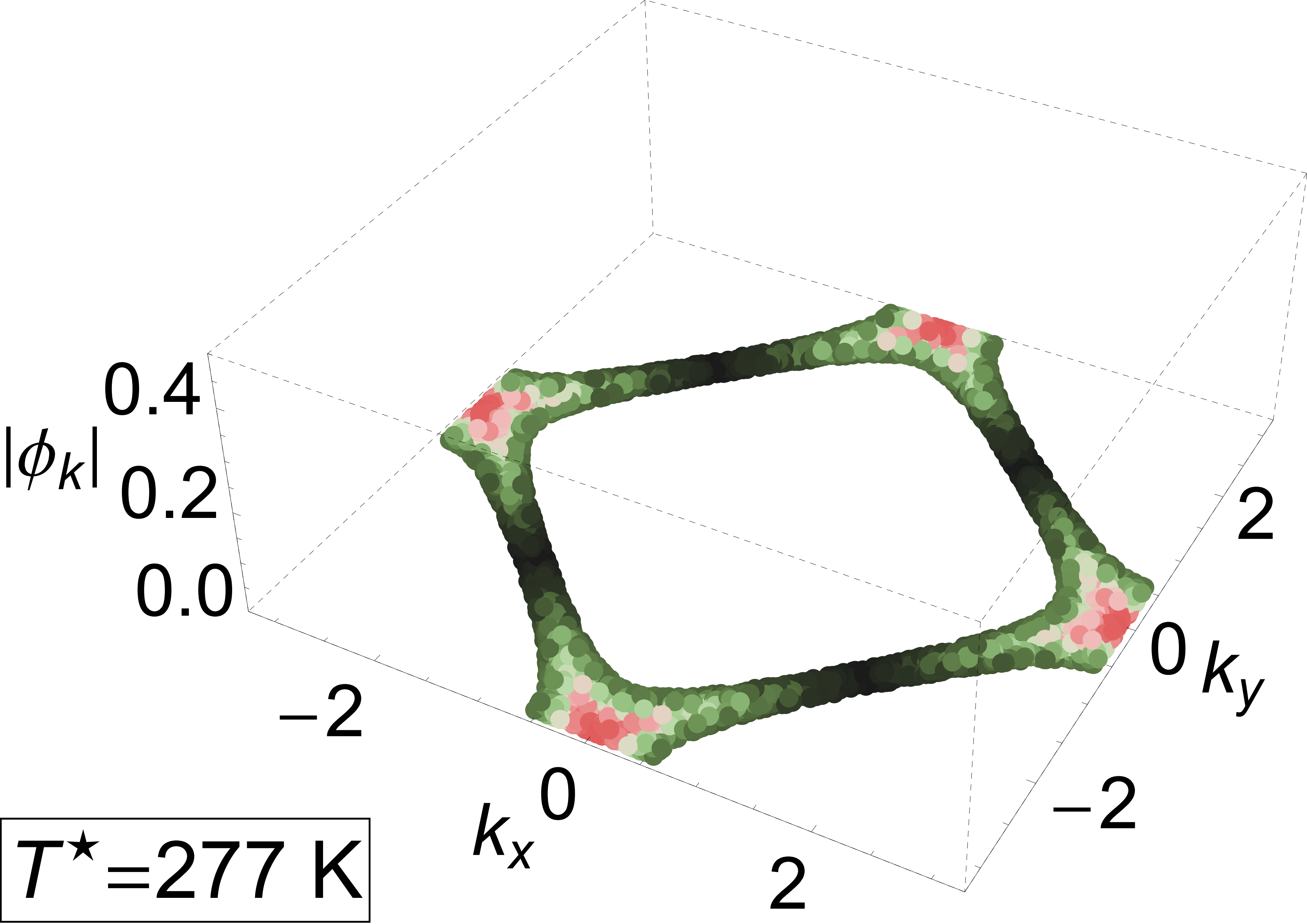}\\\vspace{0.5cm}
\includegraphics[width=0.5\columnwidth]{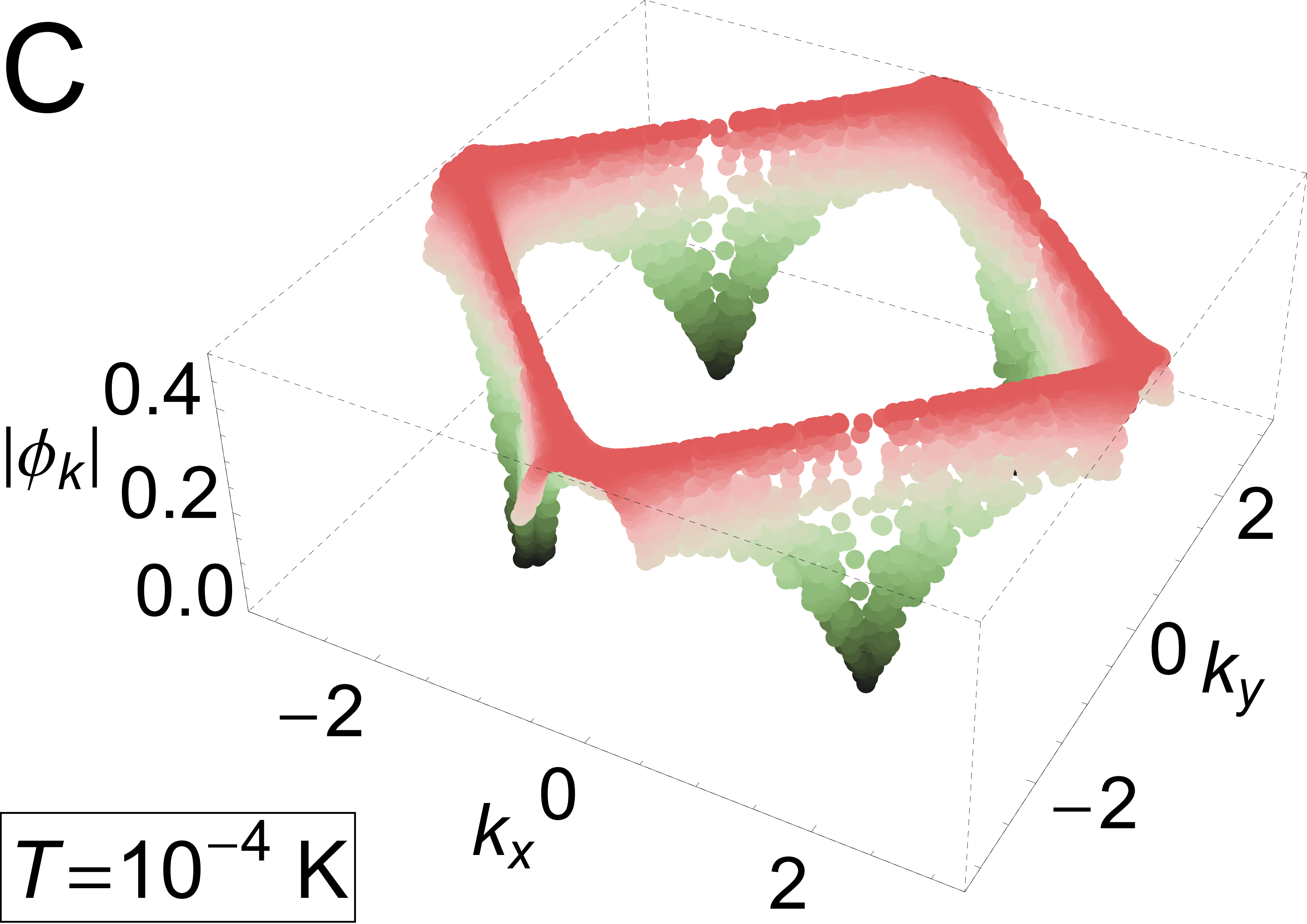}
\includegraphics[width=0.5\columnwidth]{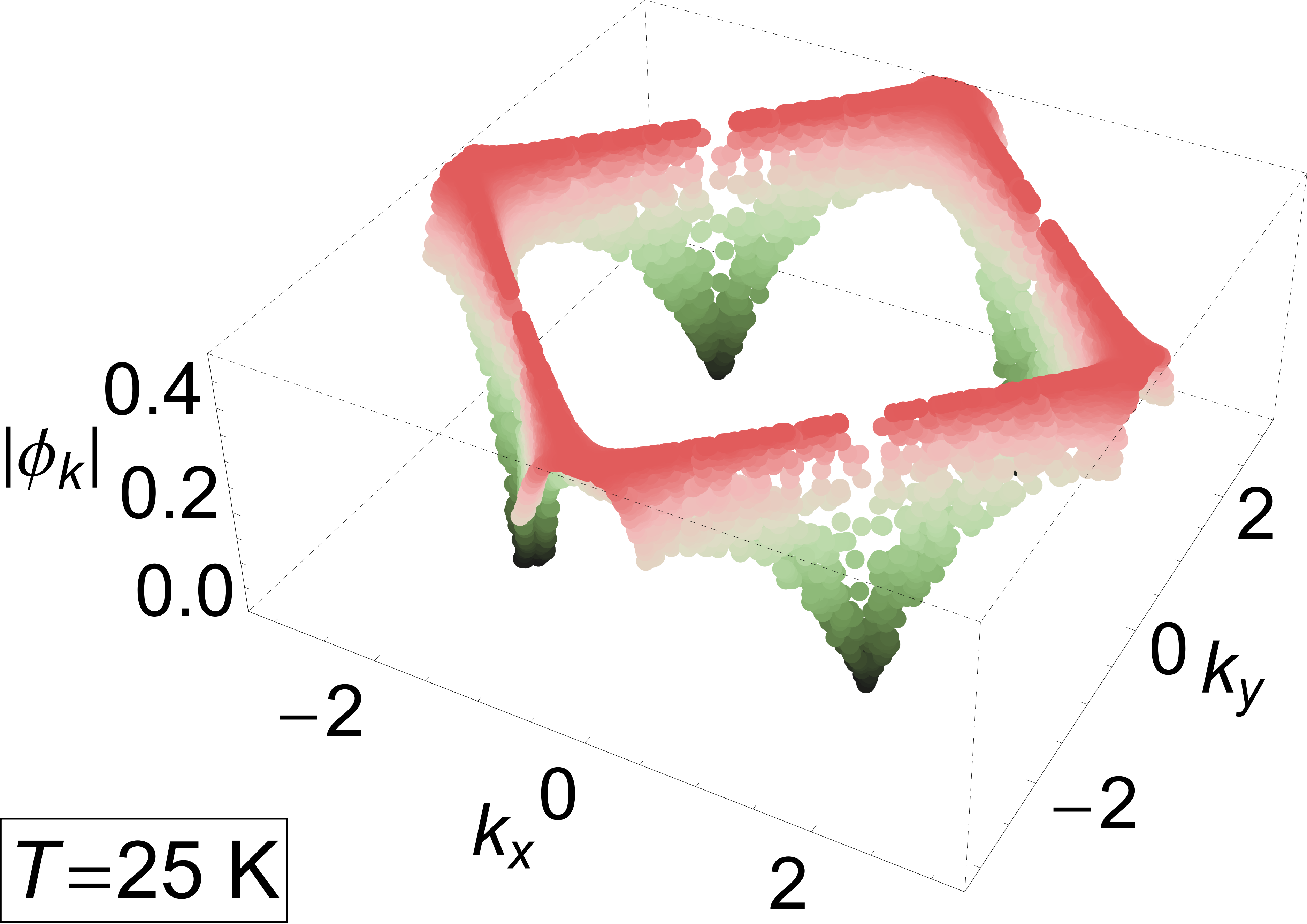}
\includegraphics[width=0.5\columnwidth]{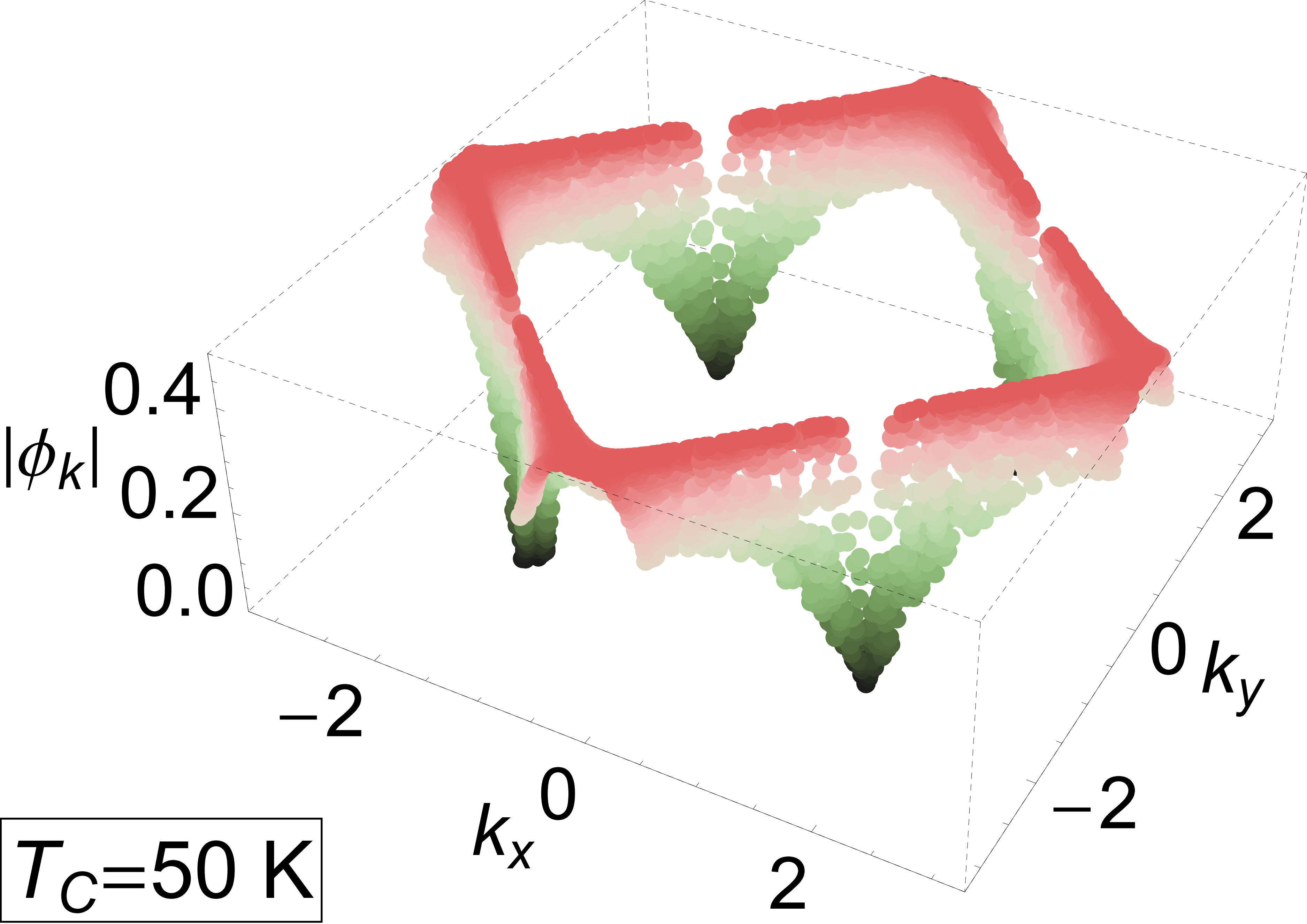}
\includegraphics[width=0.5\columnwidth]{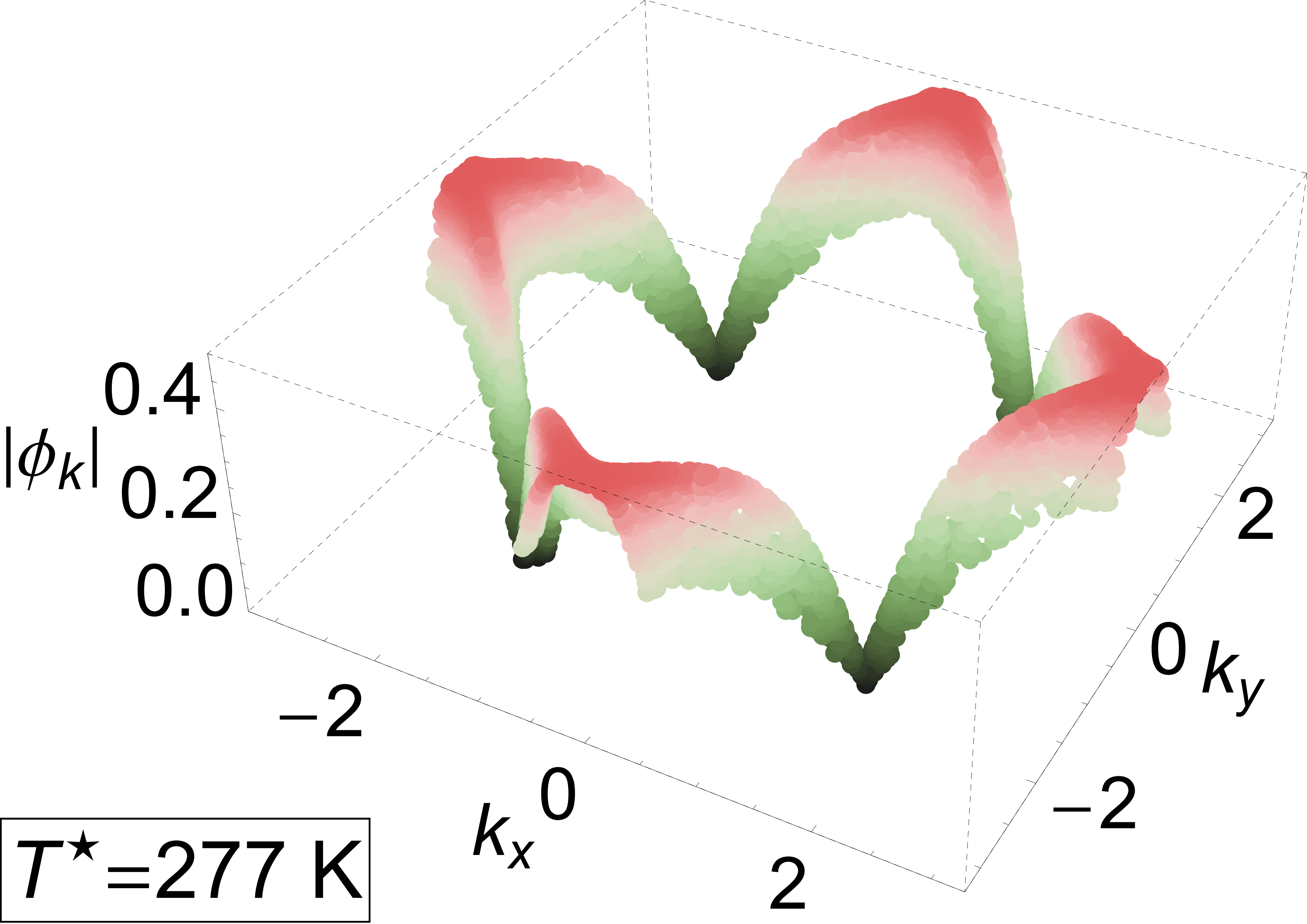}
\caption{Underdoped Bi2212 with $T_C=50$ K: the amplitude of the anomalous thermal average close to the Fermi energy for selected values of temperature and (A) $U=0$ meV, (B) $U=4$ meV, and (C) $U=12.57$ meV. \\\\}
\label{f2}
\includegraphics*[width=2\columnwidth]{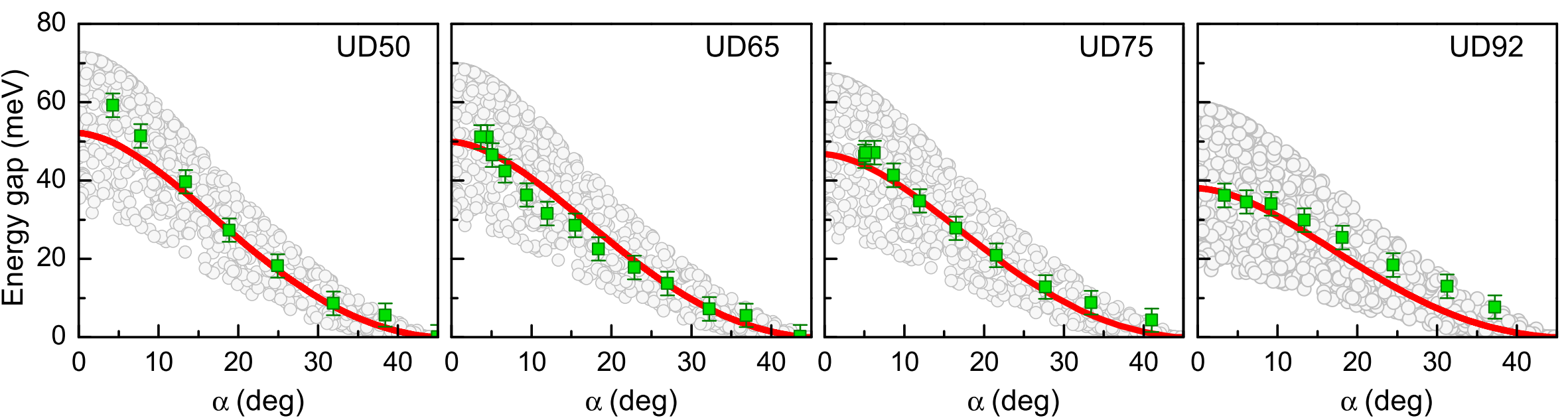}
\caption{Doping dependences of the superconducting energy gap ($T=10$ K).
The circles and red lines represent the values of the energy gap and average values of the energy gap, respectively. The green squares represent the experimental values of the energy gap taken from \cite{Vishik2010}.}
\label{f3}
\end{figure*}
%
The obtained results for selected $p$ values are collected and presented in \tab{t1}. It can be observed that with the increase of hole concentration, the critical temperature and $V$ increase, whereas the pseudogap temperature and $U$ significantly decrease. Moreover, $V$ is less than $U$ for the entire investigated range.

In \fig{f2}, we present the amplitude of the anomalous thermal average ($|\phi_{\kvec}|$) close to the Fermi energy for $p=0.08$. In particular, we take the value of $V$ determined for UD50 (underdoped, $T_C=50$ K) and different values of the EEPh interaction potential: $U\in\left\lbrace 0, 4, 12.57\right\rbrace $ meV.
We note that for $U=0$~meV (\fig{f2}A), the amplitude decreases with increasing temperature and disappears at $T_C=50$ K. 
%
\begin{figure*}[!t]
\includegraphics*[width=2\columnwidth]{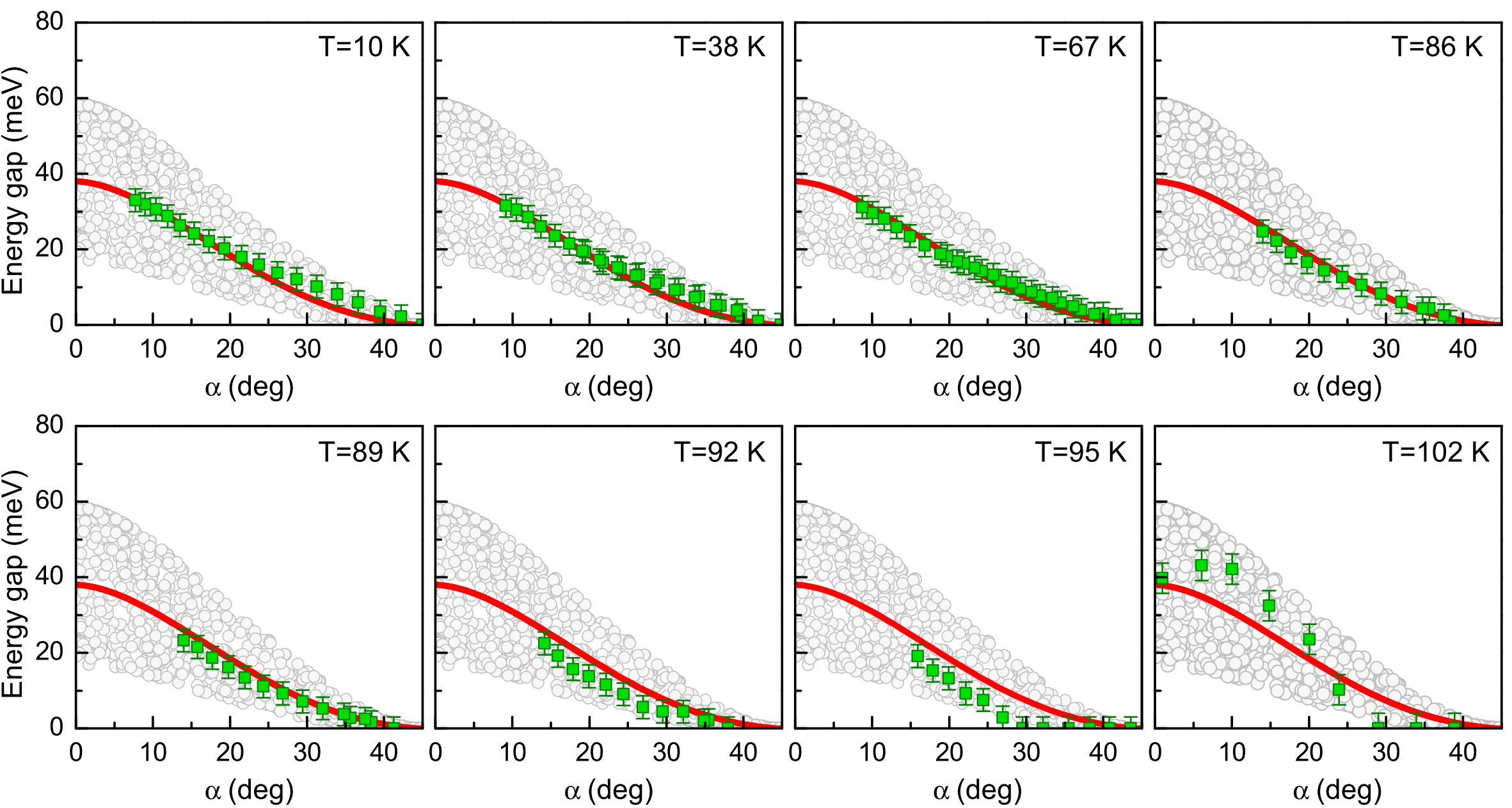}
\caption{Underdoped Bi2212 with $T_C=92$ K: the temperature dependences of the superconducting energy gap.
The circles and red lines represent the values of the energy gap and average values of the energy gap, respectively. The green squares represent the experimental values of the energy gap taken from \cite{Vishik2012} and \cite{Hashimoto2014}.}
\label{f4}
\end{figure*}
%
The gap in the antinodal regions does not vanish even with the EEPh channel completely turned off ($U=0$ meV); this is related to the fact that 
in the present model, we assumed the pairing mechanism for the high-$T_C$ superconductors to be based on the EPh and EEPh interactions. One would expect that if we take into account only the EPh interaction, then the order parameter should have an $s-$wave symmetry. However, it should be considered that in cuprates, a very strong on-site Coulomb repulsion exists between electrons having an energy of approximately $5$ eV. 
In this study, we do not explicitly take into account this repulsion. Nevertheless, we indirectly take it into account by demanding the formation of only inter-site Cooper pairs (the Wannier representation). Hence, through the transformation into the momentum representation, we obtain the $d-$wave symmetry. Thus, we can conclude that the pure EPh interaction is indirectly renormalized by the electron correlations with the requirement that only inter-site Cooper pairs exist.

In the intermediate region (\fig{f2}B), the energy gap decreases with increasing temperature, but the antinodal regions do not close completely at $T_C$.
In particular, on the basis of Eq.\eqref{r7}, it can be concluded that the value of the energy gap decreases from $18.12$ meV to $6.83$ meV when temperature increases from $10^{-4}$ K to $50$ K.

For $U=12.57$ meV (associated with $T_C=50$ K and $T^{\star}=277$ K), the antinodal region of $|\phi_{\kvec}|$ above the critical temperature is related to the anomalous normal state (see \fig{f2}C). 
This can suggest that the small near-nodal and large antinodal gaps are of completely different origins. The first one may be related to superconductivity and appears at the 
superconducting transition temperature, where the resistance vanishes. The second one may be related to another competing order parameter that exists well above $T_C$ and is not in direct correlation to the superconducting transition \cite{JZhao, Hufner}.

In order to compare our results with experimental results, it is useful to plot the superconducting energy gap as a function of the angle $\alpha\equiv{\rm arctan}\left(k_{y}/k_{x}\right)$. \fig{f3} presents the obtained values for four hole dopings (UD50, UD65, UD75, and UD92) at $10$ K.
The green squares denote the experimental data taken from \cite{Vishik2010}, the open circles represent numerical results, and the red lines represent the average value of the numerical results calculated on the basis of $1750$ points.
It can be noted that in the antinodal region ($\alpha=0$ deg), the energy-gap values clearly decrease with increasing values of the doping level. Moreover, the average numerical results are in excellent agreement with the experimental data. We obtained a similarly good agreement in the case of the $\rm Bi_{2}Sr_{2-x}La_{x}CuO_{6+\delta}$ superconductor \cite{DurajskiBSLCO}.
This indicates that the theoretical model based on the EPh and EEPh interactions is correct and can be successfully used to describe the energy gap in cuprates.

Our study was supplemented by the results of the temperature dependence of the superconducting energy gap for UD92 (\fig{f4}).
The experimental data taken from \cite{Vishik2012} and \cite{Hashimoto2014} can be reproduced for temperatures close to and below $T_C$ with very good accuracy.
We note that the energy gap near the node closes around $T_C$, whereas, in contrast to the behavior of the doping dependence, the gap magnitude near the antinodal region practically does not change with temperature
and does not diminish even above $T_C$. 
\section{Summary}
%
In the present study, we tested the model that describes the properties of the $d-$wave superconducting state in cuprates. 
On the basis of the exact numerical calculations for the Bi2212 superconductor, we showed that the energy-gap amplitude is very weakly dependent on the temperature, clearly dependent on the level of hole doping, and does not disappear above the critical temperature.
The theoretical predictions were compared with the experimental data from high-resolution angle-resolved photoemission spectroscopy. It was shown that the calculated results correctly reproduced the experimental measurements.

We expect that this work not only promotes a microscopic understanding of the mechanism of superconductivity of high-$T_C$ superconductors but also provides a guideline for future experiments.
%
%
%
\bibliography{bibliography}
%
%
\end{document}